\documentclass[aps,preprint,onecolumn,nofootinbib]{revtex4}%
\usepackage{amsfonts}
\usepackage{amsmath}
\usepackage{amssymb}
\usepackage{graphicx}%
\begin{document}
\preprint{IFT-P.037/2006}
\title{Gauge Formulation for Higher Order Gravity}
\author{R. R. Cuzinatto$^{1}$}
\email{cuzinatto@gmail.com}
\author{C. A. M. de Melo$^{1,2}$}
\email{cassius.anderson@gmail.com}
\author{L. G. Medeiros$^{1}$}
\email{leogmedeiros@gmail.com}
\author{P. J. Pompeia$^{1,3}$}
\email{pompeia@ift.unesp.br }
\affiliation{$^{1}$Instituto de F\'{\i}sica Te\'{o}rica, UNESP - S\~{a}o Paulo State University.}
\affiliation{Rua Pamplona 145, CEP 01405-900, S\~{a}o Paulo, SP, Brazil.}
\affiliation{$^{2}$Universidade Vale do Rio Verde de Tr\^{e}s Cora\c{c}\~{o}es,}
\affiliation{Av. Castelo Branco, 82 - Ch\'{a}cara das Rosas, P.O. Box 3050, CEP 37410-000,
Tr\^{e}s Cora\c{c}\~{o}es, MG, Brazil }
\affiliation{$^{3}$Comando-Geral de Tecnologia Aeroespacial, Instituto de Fomento e
Coordena\c{c}\~{a}o Industrial.}
\affiliation{Pra\c{c}a Mal. Eduardo Gomes 50, CEP 12228-901, S\~{a}o Jos\'{e} dos Campos,
SP, Brazil.}

\begin{abstract}
This work is an application of the second order gauge theory for the Lorentz
group, where a description of the gravitational interaction is obtained which
includes derivatives of the curvature. We analyze the form of the second field
strenght, $G=\partial F+fAF$, in terms of geometrical variables. All possible
independent Lagrangians constructed with quadratic contractions of $F$ and
quadratic contractions of $G$ are analyzed. The equations of motion for a
particular Lagrangian, which is analogous to Podolsky's term of his
Generalized Electrodynamics, are calculated. The static isotropic solution in
the linear approximation was found, exhibiting the regular Newtonian behaviour
at short distances as well as a meso-large distance modification.

\end{abstract}
\maketitle

\section{Introduction}

Nowadays there are many proposals to modify gravitation in order to solve
several problems as the present day accelerated expansion of the universe
\cite{Supernova,Modelos}, or to accommodate corrections of quantum nature
which arise from the classical effective backreaction of quantum matter in a
curved background \cite{Birrell}. Effective action is widely used in quantum
field theory as a powerful method of calculation. The Podolsky generalized
electrodynamics, for instance, can be viewed as an effective description of
quantum correction to the classical Maxwell Lagrangian \cite{Podolsky}.

For gravitation, usually higher orders terms are introduced by means of
Lagrangian contributions quadratic in the Riemann tensor and their
contractions \cite{R2}. This is inspired by 1-loop corrections in the
Einstein-Hilbert action in the quantized weak field approximation, or in the
equivalent Feynman construction of a spin-2 field on the flat Minkowski
background \cite{Feynman}. Besides this, at the quantum level, the $S$ matrix
for the Einstein theory is finite at one-loop level, but diverges at the
two-loop order \cite{QGDiv2Ord}, which motivates the introduction of
derivative terms in the Riemann tensor for the action \cite{delR}.

On the other hand, recently was proposed a second order construction of gauge
theories based on Utiyama's approach \cite{Utiyama}, which gives exactly the
same correction terms as in the Podolsky electrodynamics, but now arising from
the principle of local gauge invariance \cite{2ndOrd}. Therefore, a connection
between quantum corrections and gauge higher order terms in the action was
conjectured, which was proved be fulfilled also for the effective
Alekseev-Arbuzov-Baikov Lagrangian of the infrared regime of QCD \cite{Baikov}.

Here, we analyse the gauge formulation of the gravitational field based on the
framework of the second order gauge theory. The simplest gauge group is given
by the Lorentz homogeneous group in the context of a Riemannian description of
the gravitational field. Since the gauge field is given in such case by the
local spin connection, higher order in the gauge field involves naturally the
derivative of the curvature tensor. In this sense, the actual higher order
gravitational lagrangian should be constructed from invariants using the
covariant derivative of the Riemann tensor instead of the usual quadratic
terms in curvature.

The relationship between the algebraic gauge description and the geometrical
one is settled by means of the introduction of the tetrad field, and the
construction of the covariant derivatives associated with the both symmetries:
the local Lorentz and the global diffeomorphic coordinate transformations. We
use Latin indexes, $a,b,...$, for the internal Lorentz group and greek indexes
for the tangent space of the space-time manifold.

The paper is structured as follows. In section \ref{sec-Covariance} we review
some results relating gauge invariance and gravitation. The field strengths
$F$ and $G$ of the second order treatment are introduced in section
\ref{sec-GaugeFields}, where they are also written in their geometrical
counterparts: the Riemann curvature tensor and its covariant derivative.

Section \ref{sec-Counting} deals with the possible quadratic invariants of the
type $F^{2}$ and $G^{2}$. All the possible contractions are studied and only
the independent invariants are kept. This counting is made in the same spirit
as the systematic selection of the independent Riemann monomials done in
\cite{Fulling}\ and \cite{Franceses}. In the following section, section
\ref{sec-Condition}, these invariants are shown to satisfy the identity which
restricts the theories that may be called of the gauge type.

Among all invariants, we select $L_{P}=\frac{1}{2}h~\delta^{\rho}R_{\rho\chi
}\delta_{\mu}R^{\mu\chi}$, the Podolsky-like Lagrangian, for calculating the
equation of motion of the gravitational field. This higher order gravity
application is done in section \ref{sec-EqMotion}. For this Lagrangian, we
calculate the static isotropic solution in the linear regime at section
\ref{StaticIsotropicSolution}, finding the regular Newtonian potential at
short scales, but with a modified potential at intermediary scales.

Final remarks are given in section \ref{sec-Conclusion}.

\section{Gauge Interaction and Covariance\label{sec-Covariance}}

In 1956 Utiyama \cite{Utiyama} has shown how to implement a gauge description
for gravitational interaction with matter fields $Q^{A}\left(  x\right)  $
transforming according to
\begin{equation}
\delta Q^{A}(x)=\frac{1}{2}\varepsilon^{ab}(x)\left(  \Sigma_{ab}\right)
_{\,B}^{A}Q^{B}~, \label{1.1 linha}%
\end{equation}
as an implementation of the local invariance exigency of the action under
continuous proper Lorentz transformations, which are characterized by the
generators $\Sigma_{ab}$ satisfying the operation of a typical Lie group,%
\begin{equation}
\left[  \Sigma_{ab},\Sigma_{cd}\right]  =\frac{1}{2}f_{ab\,,\text{ }%
cd}^{\text{ }ef}\Sigma_{ef}~, \label{relComut}%
\end{equation}
where%
\[
\,f_{ab\,,\text{ }cd}^{\text{ }ef}=\left\{  \left[  \eta_{bc}\delta_{\,a}%
^{e}-\eta_{ac}\delta_{\,b}^{e}\right]  \delta_{\,d}^{f}-\left[  \eta
_{bd}\delta_{\,a}^{e}-\eta_{ad}\delta_{\,b}^{e}\right]  \delta_{\,c}%
^{f}\right\}  -e\leftrightarrow f
\]
are the structure constants obeying the Jacobi identity. $\varepsilon
^{ab}=-\varepsilon^{ba}$ are the parameters of the local transformation. The
capital latin indexes are for the components of the matter field.

It was clearly shown the need to introduce the compensating field $\omega
_{\mu}^{ab}\left(  x\right)  $ transforming as a connection,%
\begin{equation}
\delta\omega_{\text{ }\mu}^{ef}=\frac{1}{4}\varepsilon^{ab}\left(  x\right)
f_{ab\,,\text{ }cd}^{\text{ }ef}\omega_{\text{ }\mu}^{cd}+\partial_{\mu
}\varepsilon^{ef}\left(  x\right)  ~. \label{deltaOmega}%
\end{equation}

To ensure the covariance under coordinate transformations it was necessary to
define an space-time connection whose behaviour under infinitesimal
diffeomorphisms is%
\[
\bar{\delta}\Gamma_{\,\mu\alpha}^{\nu}=\frac{\partial\delta x^{\nu}}{\partial
x^{\lambda}}\Gamma_{\,\mu\alpha}^{\lambda}-\frac{\partial\delta x^{\lambda}%
}{\partial x^{\mu}}\Gamma_{\,\lambda\alpha}^{\nu}-\frac{\partial\delta
x^{\lambda}}{\partial x^{\alpha}}\Gamma_{\,\mu\lambda}^{\nu}-\frac
{\partial^{2}\delta x^{\nu}}{\partial x^{\mu}\partial x^{\alpha}}%
\]

The invariance of the theory implies that the compensating field must appear
through the \emph{gauge covariant derivative}
\begin{equation}
D_{\mu}Q^{A}\equiv\partial_{\mu}Q^{A}-\frac{1}{2}\omega_{\text{ }\mu}%
^{ab}\left(  \Sigma_{ab}\right)  _{\,B}^{A}Q^{B}\label{derivada
covariante}%
\end{equation}
i.e.,%
\begin{equation}
\delta D_{\mu}Q^{A}=\frac{1}{2}\varepsilon^{ab}\left(  \Sigma_{ab}\right)
_{\,B}^{A}D_{\mu}Q^{B}~,\label{CovTransfGaugeDeriv}%
\end{equation}
and the space-time connection must appear through the space-time covariant
derivative,%
\begin{equation}
\delta_{\mu}Q^{\lambda\nu}\equiv\partial_{\mu}Q^{\lambda\nu}+\Gamma
_{\,\mu\beta}^{\lambda}Q^{\beta\nu}+\Gamma_{\,\mu\alpha}^{\nu}Q^{\lambda
\alpha}~,\label{DLyraDGauge}%
\end{equation}
and the total covariant derivative:%
\begin{equation}
\nabla_{\mu}Q^{i\nu}=\partial_{\mu}Q^{i\nu}-\omega_{\text{ }\mu}^{ib}\eta
_{bk}Q^{k\nu}+\Gamma_{\,\mu\alpha}^{\nu}Q^{i\alpha}~.\label{DerivCovTotal}%
\end{equation}
This total derivative must commute with the mapping to the tangent space of
the manifold,\footnote{Note that the action of the total derivative on a
tangent space field is defined by%
\[
\nabla_{\mu}Q^{i}\equiv D_{\mu}Q^{i}~.
\]
}%
\begin{equation}
Q^{i\mu}\equiv h_{j}^{\,\mu}Q^{ij},\;Q^{ij}=h_{\,\nu}^{j}Q^{i\nu
}~,\label{TensorSpaceTensorGauge}%
\end{equation}%
\begin{equation}
\nabla_{\mu}Q^{i\nu}\equiv h_{j}^{\,\nu}\nabla_{\mu}Q^{ij}%
~,\label{DTotalDGauge}%
\end{equation}
where we have introduced the tetrad field $h$:%
\[
h_{\,\nu}^{j}h_{j}^{\,\mu}=\delta_{~\nu}^{\mu},\;\;h_{\,\nu}^{i}h_{j}^{\,\nu
}=\delta_{~j}^{i}~,\;g_{\mu\nu}=\eta_{ij}h_{\,\nu}^{i}h_{~\mu}^{j}%
~,\;\;\;\eta_{ij}=h_{i}^{~\mu}h_{j}^{~\nu}g_{\mu\nu}~,\;h=\sqrt{\det h_{\,\mu
}^{j}}=\sqrt{-g}~.
\]
The definition (\ref{DTotalDGauge}) implies the absolute parallelism of the
tetrad:%
\begin{equation}
\nabla_{\mu}h_{\,\alpha}^{j}\equiv0~,\label{AbsParTetr}%
\end{equation}
which can be solved for the compensating field,%
\[
\omega_{~i\mu}^{j}\equiv h_{i}^{\,\alpha}\left(  \delta_{\mu}h_{\,\alpha}%
^{j}\right)
\]
or for the space-time connection,%
\begin{equation}
\Gamma_{\,\mu\alpha}^{\nu}\equiv h_{j}^{\,\nu}\left(  D_{\mu}h_{\,\alpha}%
^{j}\right)  ~.\label{ConexDerGauge}%
\end{equation}

We will restrict our analysis to a symmetric space-time connection in order to
approach the Riemannian description. The extension to the Riemann-Cartan case
is quite natural, but would imply different types of invariants as admissible
Lagrangians (see discussion bellow).

\section{Gauge Field Lagrangian\label{sec-GaugeFields}}

The basic hypothesis we will assume is: the Lagrangian for the free gauge
potential depends on the field, its first and second order derivatives and
$L_{0}=L_{0}\left(  \omega_{\text{ }\mu}^{ef},\partial_{\nu}\omega_{\text{
}\mu}^{ef},\partial_{\rho}\partial_{\nu}\omega_{\text{ }\mu}^{ef}\right)
$\ obeys local invariance under (\ref{deltaOmega}). This enable us to use the
results presented elsewhere \cite{2ndOrd}\ to construct a gauge formulation
for higher order gravitation theories.

\subsection{The field strengths}

According to the work \cite{2ndOrd}, we can reexpress
\[
\delta L_{0}=\frac{1}{2}\frac{\partial L_{0}}{\partial\omega_{\text{ }\mu
}^{ef}}\delta\omega_{\text{ }\mu}^{ef}+\frac{1}{2}\frac{\partial L_{0}%
}{\partial\left(  \partial_{\nu}\omega_{\text{ }\mu}^{ef}\right)  }%
\delta\partial_{\nu}\omega_{\text{ }\mu}^{ef}+\frac{1}{2}\frac{\partial L_{0}%
}{\partial\left(  \partial_{\rho}\partial_{\nu}\omega_{\text{ }\mu}%
^{ef}\right)  }\delta\partial_{\rho}\partial_{\nu}\omega_{\text{ }\mu}%
^{ef}\equiv0~,
\]
splitting it into a set of four hierarchical equations after substituting
(\ref{deltaOmega}) and claiming the independence of the parameters
$\varepsilon^{ab}$ and their derivatives. Three of these functional equations
are used to conclude that
\begin{equation}
L_{0}=L_{0}\left(  F,G\right)  ;\;\;\;\;\;\frac{\partial L_{0}}{\partial
\omega_{\text{ }\mu}^{ab}}\equiv0~,\label{L0}%
\end{equation}
where%
\begin{equation}
F_{\,~\mu\nu}^{ab}=\partial_{\mu}\omega_{\,\nu}^{ab}-\partial_{\nu}%
\omega_{\,\mu}^{ab}-\eta_{cd}\omega_{~\ \mu}^{ac}\omega_{~\ \nu}^{db}%
+\eta_{cd}\omega_{~\ \nu}^{ac}\omega_{~\ \mu}^{db}\label{F}%
\end{equation}
and%
\begin{equation}
G_{\text{ \ }\beta\rho\sigma}^{ab}=D_{\beta}F_{\,\ \rho\sigma}^{ab}%
=\partial_{\beta}F_{\,\ \rho\sigma}^{ab}-\eta_{fd}\omega_{\text{ }\beta}%
^{af}F_{\,\ \rho\sigma}^{db}+\eta_{fd}\omega_{\text{ }\beta}^{af}%
F_{\,\ \rho\sigma}^{bd}~.\label{G}%
\end{equation}

The remaining hierarchical equation put in terms of the gauge fields $F$ and
$G$,%
\begin{equation}
\frac{\partial L_{0}}{\partial F_{\,\rho\sigma}^{ad}}f_{bc\,gh}^{\,ad}%
F_{\rho\sigma}^{gh}+\frac{\partial L_{0}}{\partial G_{\text{ }\beta\rho\sigma
}^{ad}}f_{bc\text{ }gh}^{\text{ }ad}G_{\text{ }\beta\rho\sigma}^{gh}%
\equiv0~,\label{cond_L0}%
\end{equation}
imposes restrictions upon the functional form eventually chosen for $L_{0}$.
Substituting the structure constants, this condition can be explicitly written
as%
\begin{equation}
\frac{\partial L_{0}}{\partial F_{\,\rho\sigma}^{ad}}\left[  \eta_{cg}%
\delta_{\,b}^{a}-\eta_{bg}\delta_{\,c}^{a}\right]  F_{\rho\sigma}^{gd}%
+\frac{\partial L_{0}}{\partial G_{\text{ }\beta\rho\sigma}^{ad}}\left[
\eta_{cg}\delta_{\,b}^{a}-\eta_{bg}\delta_{\,c}^{a}\right]  G_{\text{ }%
\beta\rho\sigma}^{gd}\equiv0~.\label{CondVarGauge}%
\end{equation}

\subsection{Geometrical variables}

In this section we will show how to interpret all objects and condition of the
previous sections in terms of a geometrical point of view. From
(\ref{AbsParTetr}) we read%
\[
\omega_{\text{ }\sigma}^{eg}=\eta^{gc}h_{c}^{~\alpha}\left(  \partial_{\sigma
}h_{\,\alpha}^{e}-\Gamma_{~\sigma\alpha}^{\nu}h_{~\nu}^{e}\right)
\]
and therefore the field strength $F$ is written as%
\[
F_{\,~\beta\sigma}^{eg}=\eta^{gc}h_{c}^{~\alpha}h_{~\gamma}^{e}\left[
\partial_{\sigma}\Gamma_{~\beta\alpha}^{\gamma}-\partial_{\beta}%
\Gamma_{~\sigma\alpha}^{\gamma}+\Gamma_{~\beta\alpha}^{\nu}\Gamma_{~\sigma\nu
}^{\gamma}-\Gamma_{~\sigma\alpha}^{\nu}\Gamma_{~\beta\nu}^{\gamma}\right]  ~,
\]
where we recognize the expression of the Riemann tensor \cite{Sabbata},%
\[
R_{\sigma\beta\alpha}^{~~~~~\gamma}\equiv\partial_{\sigma}\Gamma_{~\beta
\alpha}^{\gamma}-\partial_{\beta}\Gamma_{~\sigma\alpha}^{\gamma}%
+\Gamma_{~\beta\alpha}^{\nu}\Gamma_{~\sigma\nu}^{\gamma}-\Gamma_{~\sigma
\alpha}^{\nu}\Gamma_{~\beta\nu}^{\gamma},
\]
i.e.,
\begin{equation}
F_{\,~\beta\sigma}^{eg}=\eta^{gc}h_{c}^{~\alpha}h_{~\gamma}^{e}R_{\sigma
\beta\alpha}^{~~~~~\gamma}~.\label{FeR}%
\end{equation}

The easiest way to find the geometrical counterpart of $G$ is to apply the
geometrizing relations (\ref{DLyraDGauge},\ref{TensorSpaceTensorGauge}):%
\[
h_{a}^{~\mu}h_{b}^{~\nu}G_{\text{ \ }\beta\rho\sigma}^{ab}=h_{a}^{~\mu}%
h_{b}^{~\nu}D_{\beta}F_{\,\ \rho\sigma}^{ab}=\delta_{\beta}F_{\,\ \rho\sigma
}^{\mu\nu}~;
\]%
\[
F_{\,\ \rho\sigma}^{\mu\nu}=h_{a}^{~\mu}h_{b}^{~\nu}F_{\,\ \rho\sigma}^{ab}%
\]
and use (\ref{FeR}). We arrive at%
\begin{equation}
G_{\text{ \ }\beta\rho\sigma}^{ab}=h_{~\mu}^{a}h_{~\nu}^{b}g^{\nu\alpha}%
\delta_{\beta}R_{\sigma\rho\alpha}^{~~~~\mu}~, \label{GeDeltaR}%
\end{equation}
which is the most natural equation one would expect in view of the relation
(\ref{G}) between $F$ and $G$.

By means of the geometrical descriptions (\ref{FeR}) and (\ref{GeDeltaR}), we
are able to find%
\begin{align*}
\frac{\partial L_{0}}{\partial F_{\,\rho\sigma}^{ad}}  &  =\frac{\partial
L_{0}}{\partial R_{\lambda\beta\alpha}^{~~~~\gamma}}\frac{\partial
R_{\lambda\beta\alpha}^{~~~~\gamma}}{\partial F_{\,\rho\sigma}^{ad}}%
=\frac{\partial L_{0}}{\partial R_{\sigma\rho\alpha}^{~~~~\gamma}}\eta
_{bd}h_{~\alpha}^{b}h_{a}^{~\gamma},\\
\frac{\partial L_{0}}{\partial G_{\text{ }\beta\rho\sigma}^{ad}}  &
=\frac{\partial L_{0}^{\left(  4\right)  }}{\partial\left(  \delta_{\lambda
}R_{\gamma\nu\alpha}^{~~~~\mu}\right)  }\frac{\partial\left(  \delta_{\lambda
}R_{\gamma\nu\alpha}^{~~~~\mu}\right)  }{\partial G_{\text{ }\beta\rho\sigma
}^{ad}}=\frac{\partial L_{0}}{\partial\left(  \delta_{\beta}R_{\sigma
\rho\alpha}^{~~~~\mu}\right)  }g_{\alpha\omega}h_{a}^{~\mu}h_{d}^{~\omega}~.
\end{align*}

With these derivatives, the condition (\ref{cond_L0})\ for the gauge
Lagrangian is put in the form%
\begin{align}
&  \left.  \frac{\partial L_{0}}{\partial R_{\sigma\rho\beta}^{~~~~\theta}%
}\left[  \delta_{~\nu}^{\theta}g_{\gamma\lambda}-\delta_{~\gamma}^{\theta
}g_{\nu\lambda}\right]  R_{\sigma\rho\beta}^{~\ \ \ \lambda}+\right.
\label{CondL}\\
&  \left.  +\left[  \frac{\partial L_{0}}{\partial\left(  \delta_{\beta
}R_{\sigma\rho\alpha}^{~~~~~\gamma}\right)  }g_{\nu\lambda}-\frac{\partial
L_{0}}{\partial\left(  \delta_{\beta}R_{\sigma\rho\alpha}^{~~~~~\nu}\right)
}g_{\gamma\lambda}\right]  \delta_{\beta}R_{\sigma\rho\alpha}^{~\ \ \ \lambda
}\equiv0~.\right. \nonumber
\end{align}
This is a fundamental restriction upon the Lagragians tentatively proposed for
the theory, and it is quite useful in order to choose a specific suitable invariant.

\section{Quadratic Lagrangian Counting\label{sec-Counting}}

Our goal here is to determine all possible independent quadratic Lagrangians
constructed with the field strength tensors $F$ and $G$ considering their
various symmetries. By quadratic Lagrangians we mean invariants of the type
$FF$ or $GG$, but not mixed terms like $FG$\ (obviously with the proper
contraction of indices). We will also compute the linear case of the
Einstein-Hilbert Lagrangian.

\subsection{First Order Invariants\label{sec-CountingF}}

The symmetries to be considered in the construction of the invariants of the
type $FF$ are those inherited from $F$: skew-symmetry in each pair of indices:
$F_{\;\;\mu\nu}^{ab}=-F_{\;\;\mu\nu}^{ba}$ and $F_{\;\;\mu\nu}^{ab}%
=-F_{\;\;\nu\mu}^{ab}$. Besides these, there is another which is unveiled by
the geometrical form of $F$, eq. (\ref{FeR}), namely
\[
R_{\sigma\beta\alpha}^{~~~~\gamma}+R_{\beta\alpha\sigma}^{~~~~\gamma
}+R_{\alpha\sigma\beta}^{~~~~\gamma}\equiv0~,
\]
the familiar first Bianchi identity met in the context of the general relativity.

Once algebra and space-time indices can be transformed into each other by
means of a tetrad, we will consider a compact representation for $F$:%
\[
F_{\;\;\mu\nu}^{ab}\rightarrow F_{\;\;\mu\nu}^{ab}h_{c}^{~\mu}h_{d}^{~\nu
}\equiv\left(  abcd\right)  ~.
\]

Since the Lagrangians are all of the form $F^{2}$ with all allowed orders of
contractions, it is always possible to rename dummy indices in such a way that
the first $F$ will keep its indices in alphabetic order. In the table below it
follows all available permutations for the second $F$:%
\begin{equation}%
\begin{tabular}
[c]{|c|c|c|c|c|}\hline
& Fix. $a$ & Fix. $b$ & Fix. $c$ & Fix. $d$\\\hline
cyclic & $%
\begin{array}
[c]{c}%
\left(  abcd\right) \\
\left(  acdb\right) \\
\left(  adbc\right)
\end{array}
$ & $%
\begin{array}
[c]{c}%
\left(  bacd\right) \\
\left(  bcda\right) \\
\left(  bdac\right)
\end{array}
$ & $%
\begin{array}
[c]{c}%
\left(  cabd\right) \\
\left(  cbda\right) \\
\left(  cdab\right)
\end{array}
$ & $%
\begin{array}
[c]{c}%
\left(  dabc\right) \\
\left(  dbca\right) \\
\left(  dcab\right)
\end{array}
$\\\hline
non-cycl. & $%
\begin{array}
[c]{c}%
\left(  abdc\right) \\
\left(  acbd\right) \\
\left(  adcb\right)
\end{array}
$ & $%
\begin{array}
[c]{c}%
\left(  badc\right) \\
\left(  bcad\right) \\
\left(  bdca\right)
\end{array}
$ & $%
\begin{array}
[c]{c}%
\left(  cadb\right) \\
\left(  cbad\right) \\
\left(  cdba\right)
\end{array}
$ & $%
\begin{array}
[c]{c}%
\left(  dacb\right) \\
\left(  dbac\right) \\
\left(  dcba\right)
\end{array}
$\\\hline
\end{tabular}
\ \ \label{Table1}%
\end{equation}

By means of a change in one pair of indices, one can see that the non-cyclic
permutations are all proportional to the cyclic ones. Considering only the
cyclic permutations and changing two pairs of indices, the table is reduced
to:%
\[%
\begin{tabular}
[c]{|c|c|c|c|c|}\hline
& Fix. $a$ & Fix. $b$ & Fix. $c$ & Fix. $d$\\\hline
cyclical & $%
\begin{array}
[c]{c}%
\left(  abcd\right) \\
\left(  acdb\right) \\
\left(  adbc\right)
\end{array}
$ & $%
\begin{array}
[c]{c}%
-\\
\left(  bcda\right) \\
\left(  bdac\right)
\end{array}
$ & $%
\begin{array}
[c]{c}%
-\\
-\\
\left(  cdab\right)
\end{array}
$ & $%
\begin{array}
[c]{c}%
-\\
-\\
-
\end{array}
$\\\hline
\end{tabular}
\ \ \
\]

The skew-symmetries of the first $F$ (which has been taken in alphabetic
order) leads one to restrict once more the possible contractions to the three
quadratic invariants%
\begin{align}
I_{1}^{F}  &  =\left(  abcd\right)  \left(  abcd\right) \nonumber\\
I_{2}^{F}  &  =\left(  abcd\right)  \left(  acdb\right) \label{F2Inv}\\
I_{3}^{F}  &  =\left(  abcd\right)  \left(  cdab\right)  ~.\nonumber
\end{align}

We now analyze invariants constructed with one trace of $F$. The only non-null
type of trace are those obtained by contracting one index of the first pair
with one index of the second pair, in view of the skew-symmetry of this
object. All possibilities are proportional to%
\[
TrF\rightarrow h_{c}^{~\nu}F_{\;\;\mu\nu}^{ca}h_{b}^{~\mu}\equiv\left(  \cdot
ab\cdot\right)  \text{ \ or \ }\left(  \circ ab\circ\right)  ~.
\]

The quadratic invariants are given by,%
\begin{align}
I_{1}^{TrF}  &  =\left(  \cdot ab\cdot\right)  \left(  \circ ab\circ\right)
\label{TrFInv}\\
I_{2}^{TrF}  &  =\left(  \cdot ab\cdot\right)  \left(  \circ ba\circ\right)
~.\nonumber
\end{align}

Still, one can construct a linear invariant taking a double trace of $F$:%
\[
I^{TrTrF}=h_{c}^{~\nu}F_{\;\;\mu\nu}^{ca}h_{a}^{~\mu}\equiv\left(  \cdot
\circ\circ\cdot\right)  ~.
\]

\subsection{Second Order Invariants}

Let us introduce a similar notation to the one used in the case of $F$, i.e.,%
\[
G_{\text{ \ \ }\beta\rho\sigma}^{ab}\rightarrow h_{c}^{~\beta}h_{d}^{~\rho
}h_{e}^{~\sigma}G_{\text{ \ \ }\beta\rho\sigma}^{ab}\equiv\left[
abcde\right]  ~,
\]
where we identify the following symmetries:

\begin{enumerate}
\item[\textit{(i)}] antisymmetry by permutation of indices in the first pair
and last two of them,%
\[
\left[  abcde\right]  =-\left[  bacde\right]  =-\left[  abced\right]  ~;
\]

\item[\textit{(ii)}] Bianchi identity among the last three indices,%
\[
\left[  abcde\right]  +\left[  abdec\right]  +\left[  abecd\right]  =0~.
\]

\end{enumerate}

\subsubsection{Invariants of $GG$ kind}

The quadratic combinations are now in a larger amount than in the $F^{2}$
case. In fact, we have five tables like (\ref{Table1}), one to each letter
labeling, since we can associate%
\[
\left[  abcde\right]  =c\left(  abde\right)  ~.
\]

Using the symmetries cited above, one finds that the $5!$ $G^{2}$ invariants
are reduce to just two kinds:%
\begin{align*}
I_{1}^{G}  &  =\left[  abcde\right]  \left[  abcde\right] \\
I_{2}^{G}  &  =\left[  abcde\right]  \left[  debac\right]  ~.
\end{align*}
The detailed and cumbersome calculations are made in the appendix
\ref{sec-CountingG}.

\subsubsection{Invariants Involving Traces}

There are three independent types of traces for $G$:%
\begin{align}
T_{abc}^{\left(  1\right)  }  &  =h_{d}^{~\beta}G_{\;\;\beta\rho\sigma}%
^{da}h_{b}^{~\rho}h_{c}^{~\sigma}\equiv\left[  \cdot a\cdot bc\right]
\nonumber\\
T_{abc}^{\left(  2\right)  }  &  =h_{d}^{~\rho}G_{\;\;\beta\rho\sigma}%
^{da}h_{b}^{~\beta}h_{c}^{~\sigma}\equiv\left[  \cdot ab\cdot c\right]
\label{Traces}\\
T_{abc}^{\left(  3\right)  }  &  =g^{\beta\rho}G_{\;\;\beta\rho\sigma}%
^{ab}h_{c}^{~\sigma}\equiv\left[  ab\cdot\cdot c\right]  ~.\nonumber
\end{align}
Again using symmetries (see appendix \ref{sec-CountingG}) we arrive at:%
\begin{equation}%
\begin{tabular}
[c]{|c|c|}\hline
$Tr\mathcal{G}_{3}=\left[  \cdot ab\cdot c\right]  \left[  \cdot a\cdot
bc\right]  $ & $Tr\mathcal{G}_{11}=\left[  \cdot ab\cdot c\right]  \left[
\cdot bc\cdot a\right]  $\\\hline
$Tr\mathcal{G}_{5}=\left[  \cdot ab\cdot c\right]  \left[  \cdot c\cdot
ab\right]  $ & $Tr\mathcal{G}_{14}=\left[  \cdot ab\cdot c\right]  \left[
\cdot ab\cdot c\right]  $\\\hline
$Tr\mathcal{G}_{6}=\left[  ab\cdot\cdot c\right]  \left[  \cdot a\cdot
bc\right]  $ & $Tr\mathcal{G}_{17}=\left[  ab\cdot\cdot c\right]  \left[
ab\cdot\cdot c\right]  $\\\hline
$Tr\mathcal{G}_{10}=\left[  \cdot ab\cdot c\right]  \left[  \cdot ba\cdot
c\right]  $ & $Tr\mathcal{G}_{18}=\left[  ab\cdot\cdot c\right]  \left[
ac\cdot\cdot b\right]  $\\\hline
\end{tabular}
\ \ \ \ \label{TraceG}%
\end{equation}
while for double traces we have:%
\begin{align}
TrTr\mathcal{G}_{1}  &  =\left[  \cdot\circ b\cdot\circ\right]  \left[
\cdot\circ b\cdot\circ\right] \nonumber\\
TrTr\mathcal{G}_{2}  &  =\left[  \circ b\cdot\cdot\circ\right]  \left[  \circ
b\cdot\cdot\circ\right] \label{DoubleTraceG}\\
TrTr\mathcal{G}_{3}  &  =\left[  \cdot\circ b\cdot\circ\right]  \left[  \circ
b\cdot\cdot\circ\right]  ~.\nonumber
\end{align}

\subsection{Bianchi Identities}

As we already said, until now we have not used the first Bianchi identity:%
\begin{equation}
R^{\sigma\chi\rho\beta}+R^{\chi\rho\sigma\beta}+R^{\rho\sigma\chi\beta}%
\equiv0~. \label{1Bianchi}%
\end{equation}

In geometrical variables, the cyclic property of $G$ is translated to the
second Bianchi identity:%
\[
\delta^{\mu}R^{\sigma\chi\rho\beta}+\delta^{\sigma}R^{\chi\mu\rho\beta}%
+\delta^{\chi}R^{\mu\sigma\rho\beta}\equiv0~.
\]
These identities reduce the number of independent invariants, since $F\propto
R$ and $G\propto\delta R$.

\subsubsection{Reducing invariants}

Let us begin by invariants of form $F^{2}$. The first three are (\ref{F2Inv}):%
\begin{align*}
I_{1}^{F}  &  =R_{\sigma\rho\chi\kappa}R^{\sigma\rho\chi\kappa}\\
I_{2}^{F}  &  =R_{\sigma\rho\chi\kappa}R^{\sigma\chi\rho\kappa}\\
I_{3}^{F}  &  =R_{\sigma\rho\chi\kappa}R^{\chi\kappa\sigma\rho}~.
\end{align*}

As a consequence of the first Bianchi identity (\ref{1Bianchi}) and the
skewsymmetries, the curvature tensor obey:%
\begin{equation}
R_{\sigma\rho\chi\kappa}=R_{\chi\kappa\sigma\rho}~.\label{pares}%
\end{equation}
Then,%
\[
I_{3}^{F}=R_{\sigma\rho\chi\kappa}R^{\chi\kappa\sigma\rho}=R_{\sigma\rho
\chi\kappa}R^{\sigma\rho\chi\kappa}=I_{1}^{F}~,
\]
while for $I_{2}^{F}$\ one finds,%
\[
I_{2}^{F}=R_{\sigma\rho\chi\kappa}R^{\sigma\chi\rho\kappa}=-\left(
R_{\rho\chi\sigma\kappa}+R_{\chi\sigma\rho\kappa}\right)  R^{\sigma\chi
\rho\kappa}=-R_{\chi\rho\sigma\kappa}R^{\chi\sigma\rho\kappa}+R_{\chi
\sigma\rho\kappa}R^{\chi\sigma\rho\kappa}=-I_{2}^{F}+I_{1}^{F}~,
\]%
\[
2I_{2}^{F}=I_{1}^{F}~,
\]
which let us with only one invariant of this kind, $I_{1}^{F}$.

Now, we translate the trace-like invariants (\ref{TrFInv}) in a geometrical
form:%
\begin{align*}
I_{1}^{TrF}  &  =R_{\rho\mu\nu}^{\;\;\;\;\;\rho}R_{\sigma}^{\;\mu\nu\sigma}\\
I_{2}^{TrF}  &  =R_{\rho\mu\nu}^{\;\;\;\;\;\rho}R_{\sigma}^{\;\nu\mu\sigma}~.
\end{align*}
Since the Ricci tensor $R_{\mu\nu}\equiv R_{\rho\mu\nu}^{\;\;\;\;\;\rho}$ is
symmetric,\footnote{Which is a consequence of the first Bianchi identity.} we
have in fact only one invariant, $I_{1}^{TrF}=R_{\mu\nu}R^{\mu\nu}$.

At last, the only invariant of double traced form in $F$ is:%
\[
I^{TrTrF}=R~.
\]

Analogously, in view of the Bianchi identities, only four invariants of the
type $G^{2}$ remains (see appendix \ref{sec-CountingG}):
\[%
\begin{tabular}
[c]{|c|c|}\hline
$I_{1}^{G}=\delta_{\beta}R_{\sigma\rho\chi\kappa}\delta^{\beta}R^{\sigma
\rho\chi\kappa}$ & $TrTr\mathcal{G}_{2}=\delta^{\rho}R_{\rho\chi}\delta_{\mu
}R^{\mu\chi}$\\\hline
$Tr\mathcal{G}_{10}=\delta_{\beta}R_{\sigma\chi}\delta^{\chi}R^{\sigma\beta}$
& $Tr\mathcal{G}_{14}=\delta_{\beta}R_{\sigma\chi}\delta^{\beta}R^{\sigma\chi
}$\\\hline
\end{tabular}
\]

\section{Gauge Invariance Condition\label{sec-Condition}}

With the invariants constructed above we collect seven types of Lagrangians
for the gravitational field:%
\begin{equation}%
\begin{tabular}
[c]{|c|c|c|c|}\hline
Lagr. & Inv. & Gauge Form & Geom. Form\\\hline
$L_{0}^{\left(  R_{1}\right)  }$ & $\left(  I^{TrTrF}\right)  ^{n}$ & $\left(
F_{\;\;ba}^{ab}\right)  ^{n}$ & $R^{n}\;,~n=1,2$\\\hline
$L_{0}^{\left(  R_{2}\right)  }$ & $I_{1}^{TrF}$ & $F_{\;\;b\mu a}^{a}%
F_{c}^{\;b\mu c}$ & $R_{\mu\nu}R^{\mu\nu}$\\\hline
$L_{0}^{\left(  R_{3}\right)  }$ & $I_{1}^{F}$ & $F_{\;\;\mu\nu}^{ab}%
F_{ab}^{\;\;\mu\nu}$ & $R_{\alpha\beta\rho\sigma}R^{\alpha\beta\rho\sigma}%
$\\\hline
$L_{0}^{\left(  G_{1}\right)  }$ & $TrTr\mathcal{G}_{2}$ & $G_{ab\beta
}^{\;\;\;\;\beta a}G_{\text{ \ }\;\;\mu c}^{cb\mu}$ & $\delta^{\rho}%
R_{\rho\chi}\delta_{\mu}R^{\mu\chi}$\\\hline
$L_{0}^{\left(  G_{2}\right)  }$ & $Tr\mathcal{G}_{14}$ & $G_{\text{ \ \ }\mu
a\sigma}^{ab}G_{cb}^{~\;\;\mu c\sigma}$ & $\delta_{\beta}R_{\sigma\chi}%
\delta^{\beta}R^{\sigma\chi}$\\\hline
$L_{0}^{\left(  G_{3}\right)  }$ & $Tr\mathcal{G}_{10}$ & $G_{\text{
\ \ \ \ }a\sigma}^{abe}G_{deb}^{\;\;\;\;d\sigma}$ & $\delta_{\beta}%
R_{\sigma\chi}\delta^{\chi}R^{\sigma\beta}$\\\hline
$L_{0}^{\left(  G_{4}\right)  }$ & $I_{1}^{G}$ & $G_{\;\;\;\mu\nu\lambda}%
^{ab}G_{ab}^{\;\;\;\mu\nu\lambda}$ & $\delta_{\beta}R_{\sigma\rho\chi\kappa
}\delta^{\beta}R^{\sigma\rho\chi\kappa}$\\\hline
\end{tabular}
\label{LagTable}%
\end{equation}

We are considering Lagrangians only up to quadratic order in $F$ and or $G$,
which also include the linear invariant $I^{TrTrF}=R$\ and their square
$R^{2}$. Actually, one can observe that if any invariant fulfills the gauge
invariance condition, then any of its power will also do it, since this
condition is linear in the derivatives $\frac{\partial L_{0}}{\partial F}$ and
$\frac{\partial L_{0}}{\partial G}$. For instance,%
\[
L_{0}=I^{n}\;,\;\frac{\partial L_{0}}{\partial F}=nI^{n-1}\frac{\partial
I}{\partial F}~.
\]
Therefore,%
\[
\frac{\partial I}{\partial F}\left[  ...\right]  F=0\Rightarrow\frac{\partial
L_{0}}{\partial F}\left[  ...\right]  F=0~,
\]
and the same follows for $G$.

Using the skewsymmetry $\nu\leftrightarrow\gamma$\ \ of the equation
(\ref{CondL})\ and the symmetry properties of the Riemann tensor, one can
easily verify that all Lagrangians densities listed in (\ref{LagTable})
accomplish the gauge invariance condition. Then, any function of these
invariants expressible\ in a Taylor series also will fulfill the gauge
invariance condition.

\section{Equations of Motion\label{sec-EqMotion}}

Here we will concentrate our attention on the effect of the term%
\[
L_{0}^{\left(  G_{1}\right)  }=\frac{1}{2}hh_{~\sigma}^{a}h_{c}^{~\nu
}G_{ab\beta}^{\;\;\;\;\beta\sigma}G_{\text{ \ }\;\;\mu\nu}^{cb\mu}=\frac{1}%
{8}h\delta^{\rho}R\delta_{\rho}R
\]
on a gravitational theory based on the Einstein-Hilbert action plus the
$L_{0}^{\left(  G_{1}\right)  }$\ term. This Lagrangian density is equivalent,
by the Bianchi identity, to the form $\frac{1}{2}h\delta^{\rho}R_{\rho\chi
}\delta_{\mu}R^{\mu\chi}$, which is clearly analogous to Podolsky's second
order term for Electrodynamics ($L_{Podolsky}\propto\partial^{\rho}F_{\rho
\chi}\partial_{\mu}F^{\mu\chi}$). The choice of the particular Lagrangian
$L_{0}^{\left(  G_{1}\right)  }$\ is mainly motivated by this analogy. Besides
this, the $L_{0}^{\left(  G_{1}\right)  }$\ term also can be viewed as a kind
of kinetic term for the scalar curvature, what approximate such description to
the usual scalar fields. Moreover, this scalar is, up a surface term, present
in the Schwinger-DeWitt renormalized effective action for an scalar field on a
curved background \cite{effAction}. Therefore, the field theory constructed
over basis on $L_{0}^{\left(  G_{1}\right)  }$\ can be considered an effective
gravitational theory.

Taking a functional variation of the tetrad field, one finds:%
\[
h=\sqrt{-g}~,\;\;\;\delta h=\frac{1}{2}hg^{\lambda\nu}\delta g_{\lambda\nu
}=hg^{\lambda\nu}h_{\;\lambda}^{a}\eta_{ab}\delta h_{\;\nu}^{b}%
\]
and%

\begin{align*}
\delta L_{0}^{\left(  G_{1}\right)  } &  =\frac{1}{4}\partial_{\rho}\left(
h\partial^{\rho}R\delta R\right)  -\frac{1}{4}\delta R\partial_{\rho}\left(
h\partial^{\rho}R\right)  +\\
&  +\frac{1}{4}h\left[  \frac{1}{2}g^{\lambda\nu}\partial^{\rho}%
R\partial_{\rho}R-g^{\mu\nu}g^{\rho\lambda}\partial_{\mu}R\partial_{\rho
}R\right]  h_{\;\lambda}^{a}\eta_{ab}\delta h_{\;\nu}^{b}~.
\end{align*}
On calculating the equations of motion,\ we must give special attention to the
last term involving%
\begin{align*}
\delta R &  =-2R_{\mu\beta}g^{\mu\nu}g^{\beta\lambda}h_{\;\lambda}^{a}%
\eta_{ab}\delta h_{\;\nu}^{b}+\\
&  +\frac{1}{h}\partial_{\alpha}\left[  h\left(  g^{\mu\nu}\delta\Gamma
_{\mu\nu}^{\alpha}-g^{\nu\alpha}\delta\Gamma_{\nu\beta}^{\beta}\right)
\right]  ~,
\end{align*}
which will include several integration by parts. After these integrations and
some cumbersome calculations, one finds:%
\[
\delta L_{0}^{\left(  G_{1}\right)  }=\frac{1}{2}\partial_{\theta}%
\mathcal{V}^{\theta}+\frac{1}{2}h\left[  \delta_{\lambda}\delta_{\nu}\left[
\Diamond R\right]  +\frac{1}{2}\delta_{\lambda}R\delta_{\nu}R-R_{\lambda\nu
}\Diamond R-g_{\lambda\nu}\Diamond\left[  \Diamond R\right]  -\frac{1}%
{4}g_{\lambda\nu}\delta^{\rho}R\delta_{\rho}R\right]  h_{a}^{~\lambda}%
\eta^{ab}\delta h_{b}^{\;\nu}~,
\]
where%
\begin{align*}
&  \left.  \mathcal{V}^{\theta}\equiv-\frac{1}{2}\left(  g^{\mu\nu}%
\delta\Gamma_{\mu\nu}^{\theta}-g^{\nu\theta}\delta\Gamma_{\nu\beta}^{\beta
}\right)  \partial_{\rho}\left(  h\partial^{\rho}R\right)  -\frac{1}%
{2}h\partial^{\theta}R\delta R+\right.  \\
&  +\frac{1}{4}h\left(  g^{\mu\nu}g^{\alpha\beta}-g^{\nu\alpha}g^{\mu\beta
}\right)  \delta_{\alpha}\left[  \Diamond R\right]  \left(  \delta_{~\nu
}^{\theta}\delta_{~\mu}^{\lambda}\delta_{~\beta}^{\eta}+\delta_{~\mu}^{\theta
}\delta_{~\beta}^{\lambda}\delta_{~\nu}^{\eta}-\delta_{~\beta}^{\theta}%
\delta_{~\nu}^{\lambda}\delta_{~\mu}^{\eta}\right)  \delta g_{\lambda\eta}~,
\end{align*}
and%
\[
\Diamond\equiv\delta_{\beta}\delta^{\beta}%
\]
is the Laplace-Beltrami operator on the Riemannian space.

Therefore, the second order contribution to the equation of motion will be%
\begin{equation}
H_{\;\nu}^{b}\equiv h^{b\lambda}\delta_{\lambda}\delta_{\nu}\left[  \Diamond
R\right]  +\frac{1}{2}h^{b\lambda}\delta_{\lambda}R\delta_{\nu}R-R_{\;\nu}%
^{b}\Diamond R-h_{~\nu}^{b}\delta_{\beta}\delta^{\beta}\left[  \Diamond
R\right]  -\frac{1}{4}h_{~\nu}^{b}\delta^{\rho}R\delta_{\rho}R~.\label{H}%
\end{equation}
Furthermore, if we include the usual first order Einstein-Hilbert and a matter
Lagrangian densities,%
\[
S_{T}=\int d^{n}x\left(  -\frac{hR}{2\chi}-\frac{\beta}{\chi}L_{0}^{\left(
G_{1}\right)  }+h\mathcal{L}_{matter}\right)  ~,
\]
the field equations become%
\begin{equation}
G_{~\nu}^{b}+\beta H_{\;\nu}^{b}=\chi T_{~\nu}^{b}~,\label{EqEinstMod}%
\end{equation}
or in a geometrical form,%
\[
R_{\lambda\nu}-\frac{1}{2}g_{\lambda\nu}R+\beta\left[  \delta_{\lambda}%
\delta_{\nu}\left(  \Diamond R\right)  +\frac{1}{2}\delta_{\lambda}%
R\delta_{\nu}R-R_{\lambda\nu}\Diamond R-g_{\lambda\nu}\Diamond\left(  \Diamond
R\right)  -\frac{1}{4}g_{\lambda\nu}\delta^{\rho}R\delta_{\rho}R\right]  =\chi
T_{\lambda\nu}~,
\]
where $G_{~\nu}^{b}$\ is the Einstein tensor and%
\[
T_{\lambda\nu}\equiv\frac{2}{h}\frac{\delta\left(  h\mathcal{L}_{matter}%
\right)  }{\delta g^{\lambda\nu}}%
\]
is the energy-momentum tensor of the matter fields written in terms of the
metric field.

By analogy to the Alekseev-Arbuzov-Baikov \cite{Baikov}, one could expect that
the higher order terms, which can be until sixth derivative order, would be
related to infrared corrections to General Relativity, giving sensible
physical effects at large scales.

\subsection{Covariant Conservation of $T_{\lambda\nu}$}

Taking the covariant divergence of (\ref{EqEinstMod}), we have%
\[
\delta^{\nu}G_{\nu\alpha}+\beta\delta^{\nu}H_{\nu\alpha}=\chi\delta^{\nu
}T_{\nu\alpha}~.
\]
Now, from the first order case, we know that%
\[
\delta^{\nu}G_{\nu\alpha}\equiv0~.
\]
Applying the divergence to the equation (\ref{H}), one finds%
\begin{gather*}
\delta^{\nu}H_{\nu\alpha}=\delta^{\nu}\delta_{\nu}\delta_{\alpha}\Diamond
R-g_{\nu\alpha}\delta^{\nu}\Diamond\left[  \Diamond R\right]  +\frac{1}%
{2}\delta_{\nu}R\delta^{\nu}\delta_{\alpha}R+\\
+\frac{1}{2}\delta^{\nu}\delta_{\nu}R\delta_{\alpha}R-\delta^{\nu}R_{\nu
\alpha}\Diamond R-R_{\nu\alpha}\delta^{\nu}\Diamond R-\frac{1}{4}g_{\nu\alpha
}\delta^{\nu}\left(  \delta^{\rho}R\delta_{\rho}R\right)  ~.
\end{gather*}
Using the commutation relation%
\[
\left[  \delta_{\nu},\delta_{\alpha}\right]  A^{\tau}=R_{\alpha\nu
}^{\;\;\;\tau\xi}A_{\xi}%
\]
and the second Bianchi identity, we arrive at%
\[
\delta^{\nu}H_{\nu\alpha}=R_{\alpha\xi}\delta^{\xi}\Diamond R-R_{\nu\alpha
}\delta^{\nu}\Diamond R+\frac{1}{2}\delta_{\nu}R\delta^{\nu}\delta_{\alpha
}R-\frac{1}{2}\delta^{\rho}R\delta_{\alpha}\delta_{\rho}R=0~.
\]

Then, the covariant conservation of $T_{\mu\nu}$\ is established:%
\[
\delta^{\mu}\left(  G_{\mu\nu}+\beta H_{\mu\nu}\right)  \equiv0\Longrightarrow
\delta^{\mu}T_{\mu\nu}=0~,
\]
as expected from the coordinate invariance of the Lagrangian density.

\subsection{Static Isotropic Solution\label{StaticIsotropicSolution}}

In the case of an static isotropic metric,%
\[
ds^{2}=e^{\nu\left(  r\right)  }dt^{2}-e^{\lambda\left(  r\right)  }%
dr^{2}-r^{2}d\theta^{2}-r^{2}\sin^{2}\theta d\phi^{2}%
\]
in the vacuum, the equations of motion (\ref{EqEinstMod}) are reduced, in the
linear approximation, to the following coupled linear equations:%
\begin{gather*}
\nu^{\prime\prime}+\frac{2}{r}\nu^{\prime}+\\
+\beta\left(  \nu^{\left(  6\right)  }+\frac{6}{r}\nu^{\left(  5\right)
}-\frac{2}{r}\lambda^{\left(  5\right)  }-\frac{2}{r^{2}}\lambda^{\left(
4\right)  }+\frac{8}{r^{3}}\lambda^{\prime\prime\prime}-\frac{24}{r^{4}%
}\lambda^{\prime\prime}+\frac{48}{r^{5}}\lambda^{\prime}-\frac{48}{r^{6}%
}\lambda\right)  =0~,\\
\frac{1}{2}\left(  \frac{\lambda^{\prime}}{r}-2\frac{\lambda}{r^{2}}+\frac
{\nu^{\prime}}{r}-\nu^{\prime\prime}\right)  +\\
-\beta\left(  \nu^{\left(  6\right)  }+\frac{3}{r}\nu^{\left(  5\right)
}-\frac{12}{r^{2}}\nu^{\left(  4\right)  }+12\frac{\nu^{\prime\prime\prime}%
}{r^{3}}-\frac{2}{r}\lambda^{\left(  5\right)  }+\frac{4}{r^{2}}%
\lambda^{\left(  4\right)  }+8\frac{\lambda^{\prime\prime\prime}}{r^{3}%
}-48\frac{\lambda^{\prime\prime}}{r^{4}}+96\frac{\lambda^{\prime}}{r^{5}%
}-96\frac{\lambda}{r^{6}}\right)  =0~.
\end{gather*}
To solve this system, we use the Frobenius method, based on a series
expansion:%
\[
\nu\left(  r\right)  =\sum_{n}\nu_{n}r^{s+n}\;,\;\lambda\left(  r\right)
=\sum_{n}\lambda_{n}r^{s+n}~.
\]
From the first terms in the series, we find $s=-1$, and the recursion
relations above:%
\begin{align*}
\lambda_{n+4} &  =\frac{\nu_{n}\left(  n-2\right)  }{4\beta\left(  n+4\right)
\left(  n+2\right)  \left(  n+1\right)  \left(  n-\frac{1}{2}\right)  }~,\\
\nu_{n+4} &  =-\frac{\nu_{n}}{2\beta\left(  n+4\right)  \left(  n+3\right)
\left(  n+2\right)  \left(  n-\frac{1}{2}\right)  }~,
\end{align*}
such that the solution can be written as%
\[
\nu\left(  r\right)  =\sum_{m=0}^{3}\nu_{m}r^{m-1}\left(  1+\sum_{n=0}%
^{\infty}c_{nm}\right)  ~,
\]%
\[
\lambda\left(  r\right)  =\sum_{m=0}^{3}\lambda_{m}r^{m-1}-\sum_{m=0}^{3}%
\nu_{m}r^{m-1}\sum_{n=0}^{\infty}\frac{\left(  4n+m-2\right)  \left(
4n+3+m\right)  }{2\left(  4n+1+m\right)  }c_{nm}~,
\]
where $\nu_{m}$ and $\lambda_{m}$, with $m\in\left\{  0,1,2,3\right\}  $, are
the integration constants specified by the boundary conditions, and%
\[
c_{nm}\equiv\left(  -\frac{r^{4}}{2\beta}\right)  ^{n+1}\frac{\left(
m+1\right)  !}{\left(  4n+m+4\right)  !}\frac{\left(  4n+m+1\right)
!!!!}{\left(  m+1\right)  !!!!}\frac{\left(  m-\frac{9}{2}\right)
!!!!}{\left(  4n+m-\frac{1}{2}\right)  !!!!}%
\]
The notation $a!!!!$ stands for:%
\[
\left(  a+4\right)  !!!!=\left(  a+4\right)  .a!!!!~.
\]

The convergence of the series, tested by the ratio test,%
\[
\lim_{n\rightarrow\infty}\left\vert \frac{\nu_{n+1}}{\nu_{n}}\right\vert
=\left\vert \frac{r^{4}}{2\beta}\right\vert \lim_{n\rightarrow\infty
}\left\vert D_{n,m}\frac{1}{\left(  4n+m+7\right)  }\right\vert =0~,
\]%
\[
\lim_{n\rightarrow\infty}\left\vert \frac{\lambda_{n+1}}{\lambda_{n}%
}\right\vert =\left\vert \frac{r^{4}}{2\beta}\right\vert \lim_{n\rightarrow
\infty}\left\vert D_{nm}\frac{\left(  4n+m+2\right)  \left(  4n+m+1\right)
}{\left(  4n+m+5\right)  \left(  4n+m+3\right)  \left(  4n+m-2\right)
}\right\vert =0~,
\]%
\[
D_{nm}\equiv\frac{1}{\left(  4n+m+8\right)  \left(  4n+m+6\right)  \left(
4n+m+\frac{7}{2}\right)  }~,
\]
shows that both are convergent with an infinite radius of convergence.

Therefore, in the first order approximation for $\beta$, we have%
\begin{align}
\nu\left(  r\right)   &  =\frac{\nu_{0}}{r}\left(  1+\frac{1}{24\beta}%
r^{4}\right)  +\nu_{1}\left(  1-\frac{1}{60\beta}r^{4}\right)
+\label{StaticSolution}\\
&  +\nu_{2}r\left(  1-\frac{1}{360\beta}r^{4}\right)  +\nu_{3}r^{2}\left(
1-\frac{1}{1050\beta}r^{4}\right)  +\mathcal{O}\left(  \beta^{2}\right)
\nonumber\\
\lambda\left(  r\right)   &  =-\frac{\nu_{0}}{r}+\lambda_{1}+\lambda
_{2}r+\lambda_{3}r^{2}+\frac{1}{6\beta}\left(  -\frac{\nu_{0}}{4}r^{3}%
+\frac{\nu_{1}}{10}r^{4}+\frac{\nu_{2}}{60}r^{5}+\frac{\nu_{3}}{175}%
r^{6}\right)  +\mathcal{O}\left(  \beta^{2}\right)  \nonumber
\end{align}

An analysis of the solution (\ref{StaticSolution}) reveals the expected weak
field behaviour at short scales $\left(  \frac{\nu_{0}}{r}\right)  $, and
deviation from this for a mesoscale, since we are dealing only with the linear
approximation. Correspondingly, we find $\nu_{0}=2GM/c^{2}$ where $M$ is the
mass of the central body.

The remaining integration constants set scale distances where modifications of
the Newtonian behaviour appear. For instance, consider the Einstein-Hilbert
theory with cosmological constant. The static spherically symmetric solution
is%
\[
\nu\left(  r\right)  =-\lambda\left(  r\right)  =1-\frac{2GM}{c^{2}}\frac
{1}{r}-\frac{\Lambda}{3}r^{2}%
\]
where the cosmological constant sets an scale distance given by the de Sitter pseudo-radius.

Analogously, in our case, the $\nu_{1}$ constant sets a constant potential,
which can be a mean nonlocal value of the effective Lagrangian proposed,
$\nu_{2}$ sets an scale distance where a constant mean force appears, and
$\nu_{3}$ represents a gradient of force, in the same way as the cosmological
constant in the example above. Similar reasoning can be developed for the
other constants in the model.

The contribution of each constant to the net force could be fixed by requiring
that it fits \ the observational data for the tests of the gravitation. This
task deserves a careful investigation of its own and is presently under
investigation by the authors by means of the study of galaxy rotation curves,
geodesic motion, perihelion shift, gravitational lenses and redshift.

\section{Conclusion\label{sec-Conclusion}}

We have applied the second order gauge theory \cite{2ndOrd}\emph{\ }to the
local gauge theory for the homogeneous Poincar\'{e} group. It was found that
the geometrical counterparts of the usual field strength $F$\ and the second
order field strength $G=DF$\ are the Riemann tensor $R$\ and its (space-time)
covariant derivative $\delta R$. It followed the analysis of the second order
invariants composed with the geometrical entities.

We demonstrate -- employing the symmetry properties of the curvature tensor --
that the only independent Lagrangian densities for the gravitational field in
a Riemannian manifold of arbitrary dimension are the seven ones\ listed in
table (\ref{LagTable}). Linear combinations of terms proportional to powers of
$R$, as the familiar quadratic term\ in the curvature, are of first order in
the gauge potential $\omega$, therefore, in the context of the second order
gauge theory, the contributions of second order in the Lagrangian density,
which are those including second derivatives of the gauge potential, are of
type $\delta R$.

We derived equations of motion using a particularly simple\ choice for the
second order gauge Lagrangian inspired in the Podolsky's proposal for a
Generalized Electrodynamics. We found the static isotropic solution of these
equations in the linear approximation, showing that at short distances the
gravitational field behaves exactly as the Newton's law, but at meso-large
distance scales the higher order contribution dominates, exhibiting a modified potential.

In the future, we will study other solutions of these field equations,
searching for massive modes which do not violate the local gauge symmetry. Our
guide in these calculations shall be the treatment given in \cite{2ndOrd}\ to
the\ $U\left(  1\right)  $\ case, where an effective mass for the photon was
derived. To do this, one naturally must concern about the determination of the
conserved current associated with the local Lorentz symmetry and the
relationship to the global diffeomorphic invariance of the theory.

Another perspective is to apply the second order equations of motion
(\ref{EqEinstMod}) to a Friedmann-Robertson-Walker metric. The goal is to seek
for a accelerated regimes of the cosmological model arising from the higher
order terms. This proposal is now under investigation.

\bigskip

\begin{center}
\textbf{ACKNOWLEDGEMENTS}
\end{center}

R. R. C. and L. G. M.\ are grateful to Funda\c{c}\~{a}o de Amparo \`{a}
Pesquisa do Estado de S. Paulo (FAPESP) for support (grants 02/05763-8 and
02/10263-4 respectively); P. J. P. thanks to CTA staff. C. A. M. M. thanks the
hospitality of the International Centre of Theoretical Physics, under
Federation Arrangement, where part of this work was done. The authors
acknowledge Prof. B. M. Pimentel, Prof. R. Aldrovandi and Prof. A. J. Accioly
for the continuing incentive to this project, and an unknown referee for
useful comments.

\appendix{}

\section{Appendix: Counting Second Order Invariants\label{sec-CountingG}}

\subsection{Counting $GG$ Invariants}

First, let us analyse how many are the possible contractions of kind $GG$.
This is done by means of tables as in the section \ref{sec-CountingF}. The
first one is constructed fixing, for instance, the last index:%
\[%
\begin{tabular}
[c]{|c|c|c|c|c|}\hline
Fix. $e$ & Fix. $a$ & Fix. $b$ & Fix. $c$ & Fix. $d$\\\hline
cyclic & $%
\begin{array}
[c]{c}%
\left(  abcd\right) \\
\left(  acdb\right) \\
\left(  adbc\right)
\end{array}
$ & $%
\begin{array}
[c]{c}%
\left(  bacd\right) \\
\left(  bcda\right) \\
\left(  bdac\right)
\end{array}
$ & $%
\begin{array}
[c]{c}%
\left(  cabd\right) \\
\left(  cbda\right) \\
\left(  cdab\right)
\end{array}
$ & $%
\begin{array}
[c]{c}%
\left(  dabc\right) \\
\left(  dbca\right) \\
\left(  dcab\right)
\end{array}
$\\\hline
non-cycl. & $%
\begin{array}
[c]{c}%
\left(  abdc\right) \\
\left(  acbd\right) \\
\left(  adcb\right)
\end{array}
$ & $%
\begin{array}
[c]{c}%
\left(  badc\right) \\
\left(  bcad\right) \\
\left(  bdca\right)
\end{array}
$ & $%
\begin{array}
[c]{c}%
\left(  cadb\right) \\
\left(  cbad\right) \\
\left(  cdba\right)
\end{array}
$ & $%
\begin{array}
[c]{c}%
\left(  dacb\right) \\
\left(  dbac\right) \\
\left(  dcba\right)
\end{array}
$\\\hline
\end{tabular}
\]
Analogous tables result when we fix the indices $d$,$c$,$b$ and $a$. For each
table, non-cyclic permutations are equivalent to cyclic ones, giving:%

\[%
\begin{tabular}
[c]{|c|c|c|c|c|}\hline
Fix. $e$ & Fix. $a$ & Fix. $b$ & Fix. $c$ & Fix. $d$\\\hline
cyclic & $%
\begin{array}
[c]{c}%
\left(  abcd\right) \\
\left(  acdb\right) \\
\left(  adbc\right)
\end{array}
$ & $%
\begin{array}
[c]{c}%
-\\
\left(  bcda\right) \\
\left(  bdac\right)
\end{array}
$ & $%
\begin{array}
[c]{c}%
-\\
-\\
\left(  cdab\right)
\end{array}
$ & $%
\begin{array}
[c]{c}%
-\\
-\\
-
\end{array}
$\\\hline
\end{tabular}
\]
an similarly for the other four tables.

Using the cyclic permutation symmetry, one can identify elements of different
tables, reducing the number of invariants. By the skew-symmetry in the first
$G$ and renaming dummy indices, it follows:%
\[%
\begin{tabular}
[c]{|c|c|}\hline
$\mathcal{G}_{1}=\left[  abcde\right]  \left[  abcde\right]  $ &
$\mathcal{G}_{6}=\left[  abcde\right]  \left[  cdbea\right]  $\\\hline
$\mathcal{G}_{2}=\left[  abcde\right]  \left[  beacd\right]  $ &
$\mathcal{G}_{7}=\left[  abcde\right]  \left[  adbec\right]  $\\\hline
$\mathcal{G}_{3}=\left[  abcde\right]  \left[  adceb\right]  $ &
$\mathcal{G}_{8}=\left[  abcde\right]  \left[  acbde\right]  $\\\hline
$\mathcal{G}_{4}=\left[  abcde\right]  \left[  aecbd\right]  $ &
$\mathcal{G}_{9}=\left[  abcde\right]  \left[  acdeb\right]  $\\\hline
$\mathcal{G}_{5}=\left[  abcde\right]  \left[  debac\right]  $ &
$\mathcal{G}_{10}=\left[  abcde\right]  \left[  abdce\right]  $\\\hline
\end{tabular}
\]

One can further apply the cyclic permutation symmetry to the first $G$\ in
these remaining invariants and reduce even more the number of independent
quantities. Beginning with $\mathcal{G}_{10}$:%
\[
\mathcal{G}_{10}=-\left(  \left[  abdec\right]  +\left[  abecd\right]
\right)  \left[  abdce\right]  =\mathcal{G}_{1}-\mathcal{G}_{10}%
\Rightarrow2\mathcal{G}_{10}=\mathcal{G}_{1}~.
\]

On the other hand, for $\mathcal{G}_{9}$:%
\[
\mathcal{G}_{9}=-\left(  \left[  abdec\right]  +\left[  abecd\right]  \right)
\left[  acdeb\right]  =2\mathcal{G}_{4}+\mathcal{G}_{9}~\Rightarrow
\mathcal{G}_{4}=0~.
\]

Proceeding in the same way, one finds the following identities:%
\begin{align*}
&  \left.  2\mathcal{G}_{10}=\mathcal{G}_{1}%
;~\ \ \ \ \ \ \ \ \ \ \ \ \ \ \ 2\mathcal{G}_{6}=\mathcal{G}_{5}~;\right. \\
&  \left.  \mathcal{G}_{2}=\mathcal{G}_{3}=\mathcal{G}_{4}=\mathcal{G}%
_{7}=\mathcal{G}_{8}=\mathcal{G}_{9}=0~.\right.
\end{align*}
which give two independent invariants,%
\[
I_{1}^{G}=\left[  abcde\right]  \left[  abcde\right]
\;,\;\;\;\;\;\;\;\;\;I_{2}^{G}=\left[  abcde\right]  \left[  debac\right]  ~.
\]

\subsection{Counting $\left(  TrG\right)  ^{2}$\ Invariants}

Starting with the three independent traces listed in (\ref{Traces}), and
considering skew-symmetries, the possible quadratic combinations are:%
\[%
\begin{tabular}
[c]{|c|c|}\hline
$Tr\mathcal{G}_{1}=\left[  \cdot a\cdot bc\right]  \left[  \cdot a\cdot
bc\right]  $ & $Tr\mathcal{G}_{10}=\left[  \cdot ab\cdot c\right]  \left[
\cdot ba\cdot c\right]  $\\\hline
$Tr\mathcal{G}_{2}=\left[  \cdot a\cdot bc\right]  \left[  \cdot b\cdot
ac\right]  $ & $Tr\mathcal{G}_{11}=\left[  \cdot ab\cdot c\right]  \left[
\cdot bc\cdot a\right]  $\\\hline
$Tr\mathcal{G}_{3}=\left[  \cdot ab\cdot c\right]  \left[  \cdot a\cdot
bc\right]  $ & $Tr\mathcal{G}_{12}=\left[  \cdot ab\cdot c\right]  \left[
\cdot ca\cdot b\right]  $\\\hline
$Tr\mathcal{G}_{4}=\left[  \cdot ab\cdot c\right]  \left[  \cdot b\cdot
ac\right]  $ & $Tr\mathcal{G}_{13}=\left[  \cdot ab\cdot c\right]  \left[
\cdot cb\cdot a\right]  $\\\hline
$Tr\mathcal{G}_{5}=\left[  \cdot ab\cdot c\right]  \left[  \cdot c\cdot
ab\right]  $ & $Tr\mathcal{G}_{14}=\left[  \cdot ab\cdot c\right]  \left[
ab\cdot\cdot c\right]  $\\\hline
$Tr\mathcal{G}_{6}=\left[  ab\cdot\cdot c\right]  \left[  \cdot a\cdot
bc\right]  $ & $Tr\mathcal{G}_{15}=\left[  \cdot ab\cdot c\right]  \left[
ac\cdot\cdot b\right]  $\\\hline
$Tr\mathcal{G}_{7}=\left[  ab\cdot\cdot c\right]  \left[  \cdot c\cdot
ab\right]  $ & $Tr\mathcal{G}_{16}=\left[  \cdot ab\cdot c\right]  \left[
bc\cdot\cdot a\right]  $\\\hline
$Tr\mathcal{G}_{8}=\left[  \cdot ab\cdot c\right]  \left[  \cdot ab\cdot
c\right]  $ & $Tr\mathcal{G}_{17}=\left[  ab\cdot\cdot c\right]  \left[
ab\cdot\cdot c\right]  $\\\hline
$Tr\mathcal{G}_{9}=\left[  \cdot ab\cdot c\right]  \left[  \cdot ac\cdot
b\right]  $ & $Tr\mathcal{G}_{18}=\left[  ab\cdot\cdot c\right]  \left[
ac\cdot\cdot b\right]  $\\\hline
\end{tabular}
\]

The last two invariants can not be converted into any other using the
symmetries at our disposal. Each one of the preceding $Tr\mathcal{G}$ must be
analyzed, case by case, in a search for eventual interdependence.

Take, for example, the 16$^{\text{th}}$ term, and rewrite it as bellow:%
\[
Tr\mathcal{G}_{16}=-\left[  \cdot acb\cdot\right]  \left[  bc\cdot\cdot
a\right]  -\left[  \cdot a\cdot cb\right]  \left[  bc\cdot\cdot a\right]
\Rightarrow2Tr\mathcal{G}_{16}=Tr\mathcal{G}_{7}~.
\]

Repeat the reasoning for, say, the 15$^{\text{th}}$ invariant:%
\[
Tr\mathcal{G}_{15}=-\left[  \cdot acb\cdot\right]  \left[  ac\cdot\cdot
b\right]  -\left[  \cdot a\cdot cb\right]  \left[  ac\cdot\cdot b\right]
=Tr\mathcal{G}_{14}-Tr\mathcal{G}_{6}~.
\]

As soon as we perform this same check for all the above invariants, only eight
of them are kept:%
\begin{equation}%
\begin{tabular}
[c]{|c|c|}\hline
$Tr\mathcal{G}_{3}=\left[  \cdot ab\cdot c\right]  \left[  \cdot a\cdot
bc\right]  $ & $Tr\mathcal{G}_{11}=\left[  \cdot ab\cdot c\right]  \left[
\cdot bc\cdot a\right]  $\\\hline
$Tr\mathcal{G}_{5}=\left[  \cdot ab\cdot c\right]  \left[  \cdot c\cdot
ab\right]  $ & $Tr\mathcal{G}_{14}=\left[  \cdot ab\cdot c\right]  \left[
\cdot ab\cdot c\right]  $\\\hline
$Tr\mathcal{G}_{6}=\left[  ab\cdot\cdot c\right]  \left[  \cdot a\cdot
bc\right]  $ & $Tr\mathcal{G}_{17}=\left[  ab\cdot\cdot c\right]  \left[
ab\cdot\cdot c\right]  $\\\hline
$Tr\mathcal{G}_{10}=\left[  \cdot ab\cdot c\right]  \left[  \cdot ba\cdot
c\right]  $ & $Tr\mathcal{G}_{18}=\left[  ab\cdot\cdot c\right]  \left[
ac\cdot\cdot b\right]  $\\\hline
\end{tabular}
\label{TraceGAp}%
\end{equation}

\subsection{Counting $\left(  TrTrG\right)  ^{2}$ Invariants}

From $T_{abc}^{\left(  1\right)  }\equiv\left[  \cdot a\cdot bc\right]  $ one
can take a trace again:%
\[
T_{c}^{\left(  1\right)  }\equiv\left[  \cdot\circ\cdot\circ c\right]  ~.
\]

From $T_{abc}^{\left(  2\right)  }\equiv\left[  \cdot ab\cdot c\right]  $ one
finds $T_{c}\equiv\left[  \cdot\circ\circ\cdot c\right]  $ which can be
reduced to $T_{c}^{\left(  1\right)  }$\ using the $G$ skewsymmetry in the
first two indexes and changing dummy indexes. Another possible trace is
constructed from $T_{abc}^{\left(  2\right)  }$:%
\begin{equation}
T_{b}^{\left(  2\right)  }\equiv\left[  \cdot\circ b\cdot\circ\right]
~.\label{T2}%
\end{equation}
But it also is not independent of $T_{c}^{\left(  1\right)  }$:
\[
T_{b}^{\left(  2\right)  }\equiv\left[  \cdot\circ b\cdot\circ\right]
=-\left[  \cdot\circ\cdot\circ b\right]  -\left[  \cdot\circ\circ
b\cdot\right]  =-T_{b}^{\left(  1\right)  }-\left[  \circ\cdot\circ\cdot
b\right]  =-2T_{b}^{\left(  1\right)  }~.
\]
Let us set $T_{b}^{\left(  2\right)  }$\ as the independent double trace.

There are an internal double trace of $T_{abc}^{\left(  3\right)  }%
\equiv\left[  ab\cdot\cdot c\right]  $ which is independent of $T_{c}^{\left(
2\right)  }$:%
\begin{equation}
T_{b}^{\left(  3\right)  }\equiv\left[  \circ b\cdot\cdot\circ\right]  ~.
\label{T3}%
\end{equation}
The other double trace of $T_{abc}^{\left(  3\right)  }$ is,%
\[
T_{b}\equiv\left[  b\circ\cdot\cdot\circ\right]  =-T_{b}^{\left(  3\right)
}~.
\]

Then, we have the following set of independent double traces:%
\begin{align}
TrTr\mathcal{G}_{1}  &  =\left[  \cdot\circ b\cdot\circ\right]  \left[
\cdot\circ b\cdot\circ\right] \nonumber\\
TrTr\mathcal{G}_{2}  &  =\left[  \circ b\cdot\cdot\circ\right]  \left[  \circ
b\cdot\cdot\circ\right] \label{DoubleTraceGAp}\\
TrTr\mathcal{G}_{3}  &  =\left[  \cdot\circ b\cdot\circ\right]  \left[  \circ
b\cdot\cdot\circ\right]  ~.\nonumber
\end{align}

\subsection{Reducing the $G^{2}$ invariants using Bianchi
identities\label{ReduzG}}

Consider the reduction of the number of quadratic invariants in $G$ by means
of Bianchi identities. Using the geometric form, the first two invariants are:%
\[
I_{1}^{G}=\delta_{\beta}R_{\sigma\rho\chi\kappa}\delta^{\beta}R^{\sigma
\rho\chi\kappa}~,~\ \ \ \ \ \ \ \ \ I_{2}^{G}=\delta_{\beta}R_{\sigma\rho
\chi\kappa}\delta^{\chi}R^{\beta\kappa\sigma\rho}~.
\]

Applying the second Bianchi identity to $I_{2}^{G}$ we have:%
\begin{align*}
I_{2}^{G}  &  =-\delta_{\beta}R_{\sigma\rho\chi\kappa}\left(  \delta^{\chi
}R^{\kappa\beta\rho\sigma}+\delta^{\beta}R^{\chi\kappa\rho\sigma}\right)
=I_{1}^{G}-I_{2}^{G}\Rightarrow\\
&  \Rightarrow2I_{2}^{G}=I_{1}^{G}~,
\end{align*}
therefore it is sufficient to consider only $I_{1}^{G}$.

Let us analyse now the trace invariants in $G$, (\ref{TraceG}):%
\[%
\begin{tabular}
[c]{|c|c|}\hline
$Tr\mathcal{G}_{3}=\delta_{\beta}R_{\sigma\zeta\chi}^{~~~~\;\zeta}\delta_{\mu
}R^{\sigma\beta\chi\mu}$ & $Tr\mathcal{G}_{11}=\delta_{\beta}R_{\sigma\chi
}\delta^{\sigma}R^{\chi\beta}$\\\hline
$Tr\mathcal{G}_{5}=\delta_{\beta}R_{\sigma\rho\chi}^{~~~~\;\rho}\delta_{\mu
}R^{\beta\chi\sigma\mu}$ & $Tr\mathcal{G}_{14}=\delta_{\beta}R_{\sigma\chi
}\delta^{\beta}R^{\sigma\chi}$\\\hline
$Tr\mathcal{G}_{6}=\delta^{\rho}R_{\sigma\rho\chi\zeta}\delta_{\kappa
}R^{\sigma\chi\zeta\kappa}$ & $Tr\mathcal{G}_{17}=\delta^{\rho}R_{\sigma
\rho\chi\kappa}\delta_{\mu}R^{\sigma\mu\chi\kappa}$\\\hline
$Tr\mathcal{G}_{10}=\delta_{\beta}R_{\sigma\chi}\delta^{\chi}R^{\sigma\beta}$
& $Tr\mathcal{G}_{18}=\delta^{\rho}R_{\sigma\rho\chi\kappa}\delta_{\mu}%
R^{\chi\mu\sigma\kappa}$\\\hline
\end{tabular}
\]
Comparing $Tr\mathcal{G}_{10}$\ with $Tr\mathcal{G}_{11}$\ one sees that both
are the same invariant, due to the symmetry of Ricci tensor.

Using the second Bianchi identity, it follows:%
\[
Tr\mathcal{G}_{3}=\delta_{\beta}R_{\sigma\chi}g_{\mu\rho}\left(  \delta
^{\beta}R^{\rho\sigma\chi\mu}+\delta^{\sigma}R^{\beta\rho\chi\mu}\right)
=Tr\mathcal{G}_{14}-Tr\mathcal{G}_{10}%
\]
and in the same way,%
\[
Tr\mathcal{G}_{5}=Tr\mathcal{G}_{6}=-\frac{1}{2}Tr\mathcal{G}_{17}%
=-Tr\mathcal{G}_{18}=-Tr\mathcal{G}_{3}~.
\]
This shows that only $Tr\mathcal{G}_{14}$\ and $Tr\mathcal{G}_{10}$\ can be
hold independent.

We apply the same technique to the double traced invariants
(\ref{DoubleTraceG}):%
\begin{align*}
TrTr\mathcal{G}_{1} &  =\delta_{\beta}R\delta^{\beta}R\\
TrTr\mathcal{G}_{2} &  =\delta^{\rho}R_{\rho\chi}\delta_{\mu}R^{\mu\chi}\\
TrTr\mathcal{G}_{3} &  =-\delta_{\beta}R\delta_{\mu}R^{\mu\beta}~.
\end{align*}
The second Bianchi identity shows us that we have only one invariant in such
case:%
\[
TrTr\mathcal{G}_{3}=\delta_{~\zeta}^{\rho}g^{\sigma\chi}\left(  \delta
_{\sigma}R_{\beta\rho\chi}^{~~~~\;\;\zeta}+\delta_{\rho}R_{\sigma\beta\chi
}^{~~~~\;\zeta}\right)  \delta_{\mu}R^{\mu\beta}=-2TrTr\mathcal{G}_{2}%
=-\frac{1}{2}TrTr\mathcal{G}_{1}~.
\]

\end{document}